\documentclass[twocolumn,showpacs,preprintnumbers,amsmath,amssymb,aps,prb]{revtex4}
\usepackage{graphicx}
\begin{document}

\title{
Dynamic Regimes for Driven Colloidal Particles on a Periodic Substrate at Commensurate and Incommensurate Fillings 
} 
\author{
D. McDermott$^{1,2}$, J. Amelang$^{1,3}$, C. J. Olson Reichhardt$^{1}$, 
and C. Reichhardt$^{1}$ } 
\affiliation{
$^1$Theoretical Division,
Los Alamos National Laboratory, Los Alamos, New Mexico 87545 USA\\ 
$^2$ Department of Physics, University of Notre Dame, Notre Dame, Indiana 46556 USA\\
$^3$ Division of Engineering and Applied Science, California Institute of Technology, Pasadena, California 91125 USA 
} 

\date{\today}
\begin{abstract}
We examine colloidal particles driven over a periodic muffin tin substrate 
using numerical simulations. 
In the absence of a driving force 
this system exhibits a rich variety of commensurate and incommensurate static phases 
in which topological defects can form domain walls, ordered stripes, superlattices, and disordered patchy regimes as a function of the filling fraction.   
When an external drive is applied, these different static phases 
generate distinct dynamical responses. 
At incommensurate fillings the flow generally occurs in the form of 
localized pulses or solitons correlated with
the motion of the topological defect structures.
We also find dynamic transitions between different 
types of moving states that are associated with
changes in the velocity force curves,
structural transitions in the topological defect arrangements,
and modifications of the velocity distributions and particle trajectories. 
We find that the dynamic transitions between ordered and disordered flows
exhibit hysteresis, while in strongly disordered regimes there is no 
hysteresis and the velocity force curves
are smooth.  When stripe patterns are present,
transport can occur along the stripe direction rather than along the
driving direction.
Structural dynamic transitions can also occur within the pinned regimes
when the applied drive causes distortions of the interstitially pinned
particles while the particles trapped at pinning sites remain immobile.
\end{abstract}
\pacs{82.70.Dd,83.80.Hj}
\maketitle

\vskip2pc

\section{Introduction}
Non-driven systems of collectively interacting particles
typically form some type of crystal structure that is generally
triangular for repulsively interacting particles confined to two dimensions 
(2D).
In the presence of a periodic substrate,
the particles adjust their positions according to
the periodicity and strength of the substrate. 
If the 
particle lattice and substrate periodicities match,
an ordered commensurate
state forms.
Similar commensurate conditions may 
recur for higher integer 
filling factors $f = n/m$, where $n$ is the number of particles 
and $m$ is the number of substrate 
minima. 
When the 
particle lattice and substrate periodicities do not match,
the
system is incommensurate \cite{L}, and the particles may sit mostly in an
integer matching configuration with isolated spots of extra or missing
particles, or the system may break up into domains 
\cite{C,1,9}.
If an external drive is applied to
the particles, the depinning threshold and dynamic response
depend strongly on the filling fraction and whether the
configuration is commensurate or incommensurate. 
Specific examples of systems that exhibit  
commensurate-incommensurate transitions 
connected to depinning and sliding dynamics  
include vortices in type-II superconductors 
interacting with nanostructured periodic pinning arrays 
\cite{36,37,35,20,17,18}, 
vortices in Bose-Einstein condensates with a co-rotating
periodic optical trap array \cite{47}, charge density wave systems \cite{48}, 
colloidal particles 
on periodic substrates \cite{31,22,27,14,25,24,52,16,23,28}, 
charged metallic balls on patterned surfaces \cite{26}, and frictional systems 
\cite{10,6,11,5,50,19,7,38}.

One system that has recently attracted attention 
as an ideal model for studying the 
dynamics of commensurate-incommensurate behavior is
repulsively interacting colloidal particles 
confined by a periodic substrate. Experiments and computational studies have
shown that a variety of colloidal orderings can arise for 
colloidal particles interacting with one dimensional (1D) periodic substrates 
\cite{30,31,40}, 
2D periodic substrates  \cite{43,44,46,39,27,16,23,49},
2D quasiperiodic substrates \cite{8,33,12,32} 
or 2D random substrates \cite{42}. 
While these studies have provided a better
understanding of the general features of commensurate-incommensurate behaviors,
being able to {\it control} colloidal ordering and 
colloidal dynamics would lead to a variety of applications including
self-assembled structures,
particle separation, and photonic crystals.   

In recent experiments, velocity versus applied force curves were obtained
for an assembly of colloidal particles 
driven over a periodic substrate \cite{2}.
These experiments showed that the largest depinning threshold occurs for
the commensurate state at which there is one colloid per substrate minimum.
At incommensurate fillings, 
the depinning threshold is substantially reduced
and the flow
occurs through localized soliton type pulse motions, where individual
particles do not move continuously through the sample but instead jump by
one lattice constant each time the soliton or kink moves past.
Just above the first matching filling, 
a forward-moving kink forms, 
while just below the first matching filling, there is
a backward-moving anti-kink or vacancy.  In both cases
the net motion of the particles is in the forward direction.
At the commensurate filling, a kink-antikink pair can be created, 
so that even for commensurate states the motion
can consist of 
temporarily mobile particles co-existing with pinned particles. 
Subsequent numerical simulations    
of Yukawa particles interacting with egg-carton 
substrates also produced kink and antikink motion, which the authors
related to an effective friction 
\cite{23,49}. 
These works showed that decreasing the substrate strength at an
incommensurate filling
leads to a 2D Aubry transition \cite{4}, where the depinning threshold 
vanishes and the colloidal lattice essentially floats on the substrate. 
In addition, 
the velocity force curves
exhibited signatures of two step depinning processes 
corresponding to the different dynamical states of soliton depinning and
the depinning of all of the particles.
Soliton-type flow for driven particle systems has 
also been studied for 
friction models \cite{6,11,5} as well as for 
vortices in
type-II superconductors interacting with periodic pinning arrays, where
the soliton flow produces distinct features in the current-voltage curves that
are analogs to the velocity-force curves in the driven colloidal systems
\cite{15,20,18,17}. 
In numerical \cite{12} and experimental \cite{41} studies of
the driven dynamics of colloidal particles moving over periodic substrates,
different dynamical locking regimes and dynamical ordering were
observed.  Here, the type of motion and amount of order in the flow
depends on the orientation of the driving direction with respect to the
symmetry directions of the substrate.
Kink motion
has also been found in experiments on colloidal particles moving over
quasiperiodic substrates \cite{8}. 
In addition, in studies of colloidal 
particles driven over corrugated 1D arrays, different dynamical transitions
between pinned, smectic and disordered flowing regimes 
were observed \cite{40}. 

The recent experiments on the dynamics of colloidal particles sliding 
over periodic arrays
focused only on fillings very near the first matching condition \cite{2}.
For higher fillings or other types of periodic substrates, 
an additional variety of dynamical phases should be expected to appear.
In a recent study \cite{1}
we described the pinned
configurations for colloidal particles 
interacting with a 2D square periodic array of pinning sites 
for fillings ranging from $f=1.0$ to $9.0$. 
We particularly focused on the regime between
$f=4.0$ and $f=5.0$ 
since it contains a transition from a
triangular colloidal lattice at $f = 4.0$ to a square lattice 
at $f = 5.0$. 
The system forms domain walls composed of topological defects 
for incommensurate fillings just above and below $f = 4.0$, 
transitions to stripe patterns 
for $4.1 < f < 4.6$, and exhibits 
disordered patchy domains for $4.6 < f < 5.0$. 
The periodic substrate in Ref.~\cite{1} is not an egg carton potential
of the type used in the recent driven colloidal experiments \cite{2} 
or in static ordering studies for colloidal particles in 2D 
arrays \cite{22,27,16}, but is instead a muffin tin structure formed from
a periodic array of localized pinning sites with small radii,
so that beyond $f = 1.0$ the additional colloidal particles 
sit in the interstitial regions between the 
particles that are trapped directly by the pinning sites.
Periodic pinning arrangements of this type have
been experimentally realized in colloidal
systems and studied in the range $f = 4.0$ to $f = 5.5$, with
a triangular colloidal lattice observed at $f = 4.0$ 
when the pinning strength was such that 
each pinning site captures only one particle \cite{34}. 
Such pinning arrays
have also been used to study
dynamical locking for colloidal particles driven over a 
substrate at different angles
with respect to the substrate symmetry directions \cite{43}. 
Experimental and numerical studies of 
vortex systems using similar 2D periodic pinning arrays
also obtained a triangular vortex lattice at $f=4.0$ 
and a square lattice at $f= 5.0$ \cite{36,35} with domain wall and stripe
states between these fillings \cite{36}. Since the static phases have 
already been characterized, colloidal particles interacting
with a muffin tin potential may be an ideal system for understanding how 
depinning dynamics occurs in the presence of domain walls 
or stripes, or in systems with coexisting ordered and disordered phases. 

\section{Simulation and System}   
Our system consists of a 
2D assembly of 
colloidal particles that interact repulsively
via a Yukawa potential.
The initial configurations of the $N_{c}$ particles 
are obtained using a simulated annealing procedure. 
After annealing, we apply an external drive 
${\bf F}^D=F_d{\bf \hat x}$ to each particle.
The system has periodic boundary conditions 
in both the $x$ and $y$ directions. 
The dynamics of the particles is obtained by 
using an overdamped equation of motion,
as in previous numerical studies \cite{22,12,16,1}. 
The motion of a single particle $i$ is given by integrating the following
equation:
\begin{equation}  
\eta \frac{d {\bf R}_{i}}{dt} = 
-\sum_{i\ne j}^{N_{i}}{\bf \nabla}V(R_{ij}) +  {\bf F}^{P}_{i} +  {\bf F}^{D} +  {\bf F}^{T}_{i}. 
\end{equation} 
Here $\eta$ is the damping constant  
and the particle-particle interaction potential is 
$V(R_{ij}) =  q^2E_{0}\exp(-\kappa R_{ij})/R_{ij}$,   
where $E_{0} = Z^{*2}/4\pi\epsilon\epsilon_{0}a_{0}$,  
$q$ is the dimensionless interaction strength,
$Z^{*}$ is the 
effective charge of the colloid,
and $\epsilon$ is the solvent dielectric constant. 
The screening length is $1/\kappa$ and the  
lengths are measured in units of $a_{0}$, time in units of 
$\tau = \eta/E_{0}$, and forces in units of $F_{0} = E_{0}/a_{0}$.
${\bf R}_{i(j)}$ is the position of particle $i(j)$,
and $R_{ij} = |{\bf R}_{i} - {\bf R}_{j}|$.
The substrate is modeled as $N_{p}$ pinning sites placed in a square array with 
a lattice constant $a$.  Each pinning site is represented by a 
parabolic potential trap 
with a radius $R_{p}=0.3a_0$ which gives rise to a pinning force of 
${\bf F}^{P}_{i} = \sum_{k=1}^{N_p}F_{p}(R_{ik}/R_{p})\Theta(R_{p}- R_{ik}){\bf \hat{R}}_{ik}$, 
where $R_{ik} = |{\bf R}_{i} - {\bf R}_{k}|$ is the distance between particle
$i$ and the center of pinning site $k$, and
${\bf {\hat R}}_{ik} = ({\bf R}_{i} - {\bf R}_{k})/R_{ik}$. 
$F_{p}=10$ is the maximum force of the 
pinning site, and $\Theta$ is the Heaviside step function.   
The thermal fluctuations come from the 
Langevin noise term $F^{T}$ with the properties
$\langle F^{T}(t)\rangle = 0$ and 
$\langle F^{T}_{i}(t)F_j^T(t^{\prime})\rangle = 2\eta k_{B}T\delta_{ij}\delta(t - t^{\prime})$,   
where $k_{B}$ is the Boltzmann constant. 

After annealing, we set $T=0$ and begin applying a drive by starting at
$F_{d} = 0$ and increasing $F_d$
in steps of $\delta F_{d} = 0.001$, 
waiting $5 \times 10^4$ simulation time steps 
at each force increment.  
We have tested the results for slower drive sweep rates 
and find that they do not change.  
We measure the average particle velocity 
$\langle V_{x}\rangle = N_c^{-1}\sum^{N_{c}}_{i = 1} {\bf v}_{i}\cdot {\hat {\bf x}} $, 
and 
$\langle V_{y}\rangle = \sum^{N_{c}}_{i = 1} {\bf v}_{i}\cdot {\hat {\bf y}}$. 
We also measure the distributions $P(v_x)$  and $P(v_y)$ of the instantaneous 
particle velocities in the $x$ and $y$ directions, respectively.
The fraction of $n-$fold coordinated particles $P_n$ is obtained using a
Voronoi construction, with $P_n=\sum_i^{N_c}\delta(n-z_i)$, where $z_i$ is
the coordination number of particle $i$.
The filling fraction is 
defined as  $f =N_{c}/N_{p}$. In this study we consider the case where each  
pinning site captures at most one particle and 
limit our range of $F_{d}$ such that
particles trapped at pinning sites do not depin, so that we can concentrate
on the initial flow of the particles trapped in the interstitial
regions.   

\begin{figure}
\includegraphics[width=3.5in]{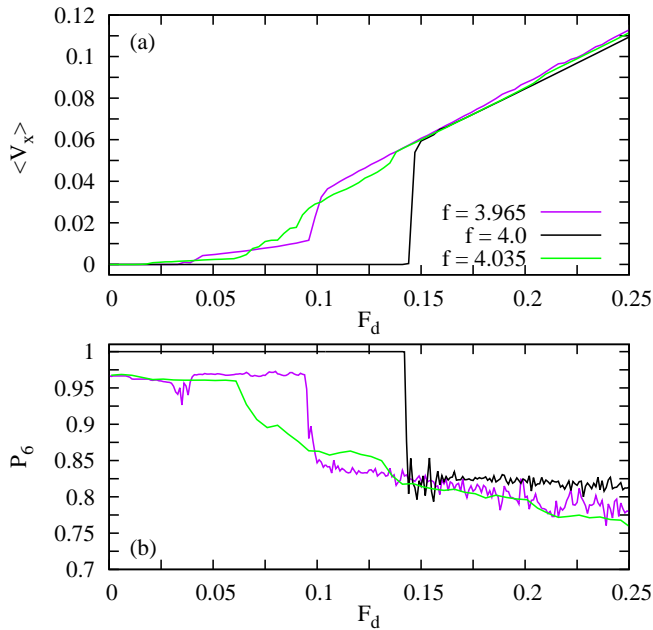}
\caption{ (a) The average velocity $\langle V_{x}\rangle$ 
vs $F_{d}$ for 
a filling of $f = 4.0$ (black),
$f = 3.965$ (purple), and $f = 4.035$ (green). 
(b) The corresponding fraction of six-fold coordinated particles
$P_{6}$ vs $F_{d}$. 
}
\label{fig:1}
\end{figure}

\section{Fillings Near $f = 4.0$} 

At $f = 4.0$, we previously found that the non-driven 
colloidal configuration is a triangular 
lattice, while for $3.9 < f < 4.0$ 
and $4.0 < f < 4.1$, 
the same triangular lattice is
interspersed with grain boundaries of 
5-7 paired dislocations separating regions where the particle lattice 
is rotated with respect to the
substrate \cite{1}. 
In Fig.~\ref{fig:1}(a) 
we plot $\langle V_{x}\rangle$ versus $F_d$ for samples with
$f = 3.965$, $4.0$,
and $4.035$, and in Fig.~1(b) we plot the corresponding fraction of
sixfold coordinated particles $P_{6}$. 
At the commensurate filling of $f = 4.0$, 
there is a finite depinning threshold near $F_{d} = 0.14$,
while for the incommensurate fillings, the depinning threshold is almost zero
for $f= 4.035$ and slightly higher than zero for
$f = 3.965$.
The velocity force curves at the incommensurate fillings 
show a multiple depinning response as
indicated by the multiple jumps in $\langle V_{x}\rangle$. 
In Fig.~\ref{fig:1}(b), $P_{6} = 1.0$ for the 
$f = 4.0$ system for drives below the depinning threshold
due to the triangular
ordering of the pinned state, but $P_6$ drops at depinning when
only the interstitially pinned particles are able to move, causing
some non-sixfold ordering 
to appear.
For the $f = 4.035$ sample, initially $P_6<1.0$
since 
the particles in the grain boundaries are $5-$ or $7-$fold coordinated, and   
for $F_{d} > 0.06$,
$P_{6}$ drops further since more non-sixfold coordinated defects are generated 
as $F_{d}$ increases. 

\begin{figure}
\includegraphics[width=3.5in]{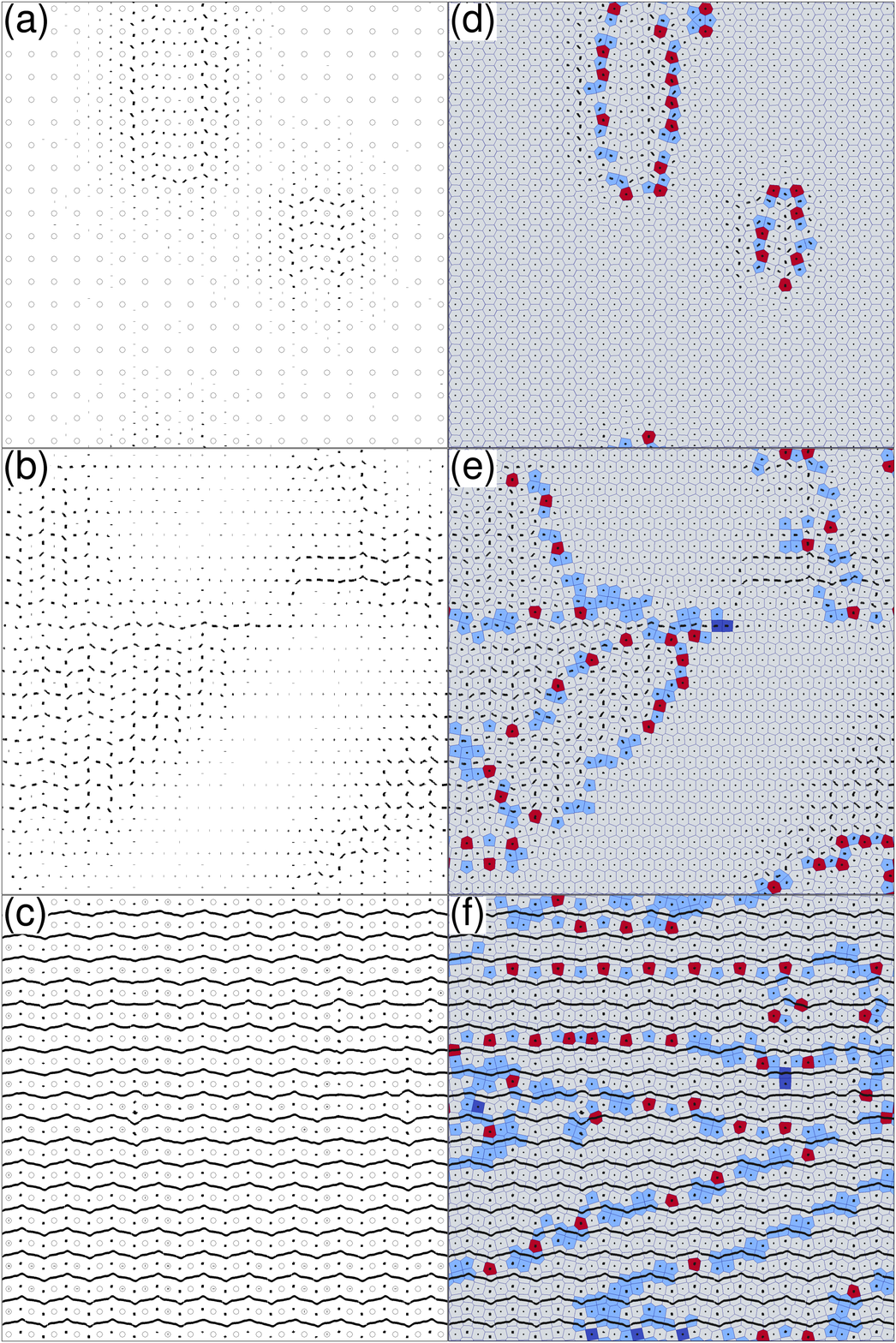}
\caption{(a-c)
The pinning site locations (open circles) and particle trajectories 
(black lines) over a fixed interval of time for the
system at $f = 4.035$. 
(a) $F_{d} = 0.05$, (b) $F_{d} = 0.08$, and (c) $F_{d}= 0.2$. 
(d-f) Pinning site locations (open circles), particle trajectories (black 
lines), and Voronoi construction (blue lines) 
for the same sample at
(d) $F_{d} = 0.05$, (e) $F_{d} = 0.08$, and (f) $F_{d} = 0.2$.
The Voronoi polygons are colored according to their number of sides: 
4 (dark blue), 5 (light blue), 6 (white), and 7 (red).
(a,d) The moving grain boundary state; (b,e) the fluctuating grain boundary
state; and (c,f) the continuous flow regime.
}
\label{fig:2}
\end{figure}

It is the depinning of the grain boundaries that is responsible for reducing
the depinning force at the incommensurate fillings.
Above depinning at these fillings, the grain boundaries move continuously but
the particles only move about a lattice constant each time
a grain boundary passes over them. 
This is illustrated in Fig.~\ref{fig:2}(a) where we plot
the particle trajectories over a fixed time
for a sample with $f = 4.035$ at $F_{d} = 0.05$. 
The particle motion is 
localized and correlated with the positions of the grain boundaries, as shown
by the corresponding Voronoi construction in Fig.~\ref{fig:2}(d);
motion occurs only where the grain boundaries are located, while regions
without grain boundaries remain pinned.
When the particles move, their trajectories are not strictly 
1D but have a zig-zag shape 
due to the fact that the particles
must shift in both the $x$ and $y$ directions to permit the grain boundary to
pass, since the particles change from one triangular lattice orientation
to the other after the grain boundary has moved over them.
This motion has similarities
to the soliton or kink motion observed near 
the $f =1.0$ filling in experiments and simulations 
\cite{2,23}, but instead of isolated 
moving kinks or antikinks, we have a moving 
row of correlated kinks or anti-kinks that is bound to the grain boundary.
The shape of the moving domain that the grain boundary defines does not
remain completely static
but can undergo expansions and contractions as it moves. 

For $F_d = 0.08$, we find a transition from well-defined domain walls 
to a state where the domain walls break apart and reform, as illustrated
in Fig.~\ref{fig:2}(b).  The motion still remains localized to the vicinity
of the domain walls, as shown in Fig.~\ref{fig:2}(e).
The domain wall number strongly fluctuates in this state, which
resembles the kink-antikink nucleation regime observed in 
the experiments of Ref.~\cite{2}.   
At $F_{d} = 0.2$, the flow of the mobile particles is continuous, but a 
portion of
the interstitially pinned particles, as well as all of the directly pinned
particles, remain immobile.
In this regime, the soliton motion is lost, as illustrated in 
Fig.~\ref{fig:2}(c).  
In addition, the grain boundary state is destroyed due to a
proliferation of 5-fold and 7-fold coordinated particles, as shown in
Fig.~\ref{fig:2}(f).  
At $f=3.965$, where there is also a grain boundary pinned state, 
we find a sequence of dynamical states very similar to those shown 
in Fig.~\ref{fig:2} for
$f = 4.035$. 

\begin{figure*}
\includegraphics[width=5.0in]{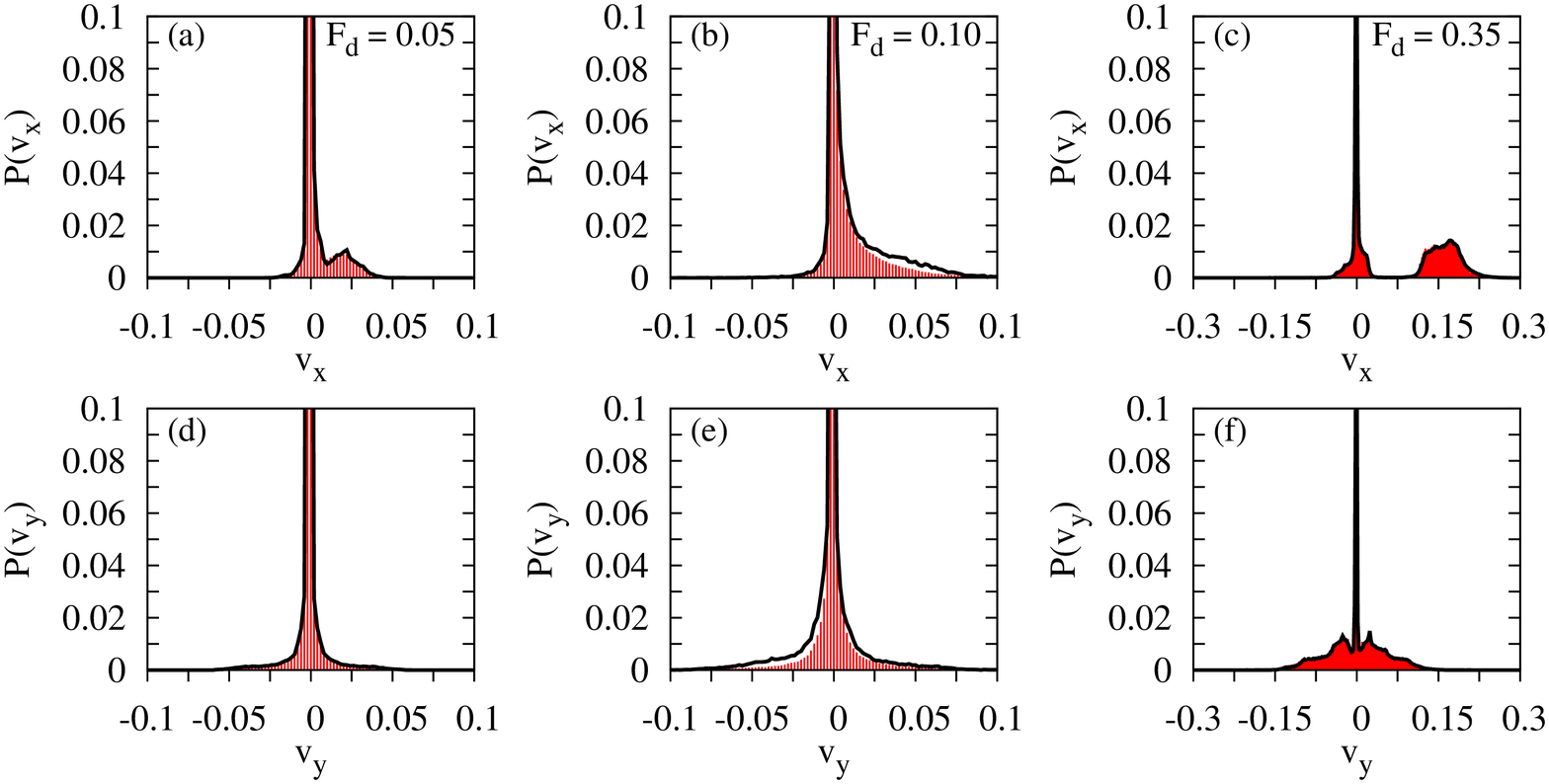}
\caption{The velocity histograms for the system in Fig.~\ref{fig:2}
with $f=4.035$.
(a,b,c) $P(v_{x})$; (d,e,f) $P(v_{y})$.  (a,d) $F_{d} = 0.05$. 
(b,e) $F_{d} = 0.8$. (c,f) $F_{d} = 0.2$.  Black lines: increasing
sweep of $F_d$.  Filled red curves: decreasing sweep of $F_d$. 
}
\label{fig:3}
\end{figure*}

We can identify 
four distinct dynamical regimes for the grain boundary forming fillings:
(I) pinned; (II) grain boundary motion; (III) moving fluctuating grain
boundaries; and (IV) continuous interstitial flow.
The different regimes
can also be characterized with the distributions $P(v_x)$ and $P(v_y)$
of the instantaneous particle velocities,
as shown for the 
$f = 4.035$ sample
at $F_{d} = 0.05$ in Fig.~\ref{fig:3}(a). 
Here, in the moving grain boundary state, 
there is a peak in $P(v_{x})$ at $v_{x} = 0.0$ from the immobile particles,
a second peak near $v_{x} = 0.025$, and continuous weight between the two
peaks. 
Although the grain boundaries move at constant speed, the individual particles
start with $v_x=0$ in the pinned state, reach a maximum velocity as the grain
boundary moves past, and then drop back to the $v_x=0$ pinned state.
As a result, we observe the full range of velocities 
between $v_x=0$ and $v_x=0.025$.
Fig.~\ref{fig:3}(d) shows that the $y$ velocities of the particles at the
same drive 
are not strictly zero due to the zig-zag motion associated with the
grain boundaries; however, the average value of $v_y$ is zero.
For $F_{d} = 0.08$ in the same system, Fig.~\ref{fig:3}(b) shows that
$P(v_x)$ is skewed out to higher positive $v_x$ values than for
the $F_d=0.05$ case, but the secondary peak in $P(v_x)$ has disappeared
due to the strongly fluctuating nature of the grain boundary motion in this
regime.
In the 
corresponding $P(v_{y})$ shown in Fig.~\ref{fig:3}(e), the tails have moved
out further in the positive and negative $v_y$ directions.
For $F_{d} = 0.20$ in the continuous 
flow regime, $P(v_x)$ in Fig.~\ref{fig:3}(c) has two distinct peaks with
one centered
at $v_{x} = 0.0$ and the other centered at $v_{x} = 0.16$. 
A region of zero weight exists
between the two peaks, indicating that
there are no particles moving at these intermediate velocities. 
This is due to the loss of the soliton-like motion; in the continuous
flow regime, the mobile particles always remain in motion while the
pinned particles always remain immobile.
The corresponding $P(v_{y})$ in Fig.~\ref{fig:3}(f) is still centered 
at zero but now shows additional structure with
two satellite peaks that arise due to the
sinuous motion 
in this regime, as illustrated in Fig.~\ref{fig:2}(c).
For increasing $F_d$, $P(v_x)$ and $P(v_y)$ retain the same form shown in
Fig. 3(c,f) until the particles at the pinning sites depin, upon which a
new set of dynamical states appears that will not be considered in this work.

To test for hysteretic effects, we measure $P(v_x)$ and $P(v_y)$ both for
increasing $F_d$ up to its maximum value (black curves in Fig.~\ref{fig:3})
and for decreasing $F_d$ back down to zero (filled red curves in 
Fig.~\ref{fig:3}).
In the continuous flow regime, 
Fig.~\ref{fig:3}(c) shows that 
there is no hysteresis; however, in the moving fluctuating grain boundary
regime  [Fig.~\ref{fig:3}(b)]
the system is generally
more disordered when $F_d$ is being decreased,
indicating that the fluctuating grain boundary regime
persists down to a lower drive on the downward sweep of $F_d$
than the drive at which it first appeared
during the upward sweep of $F_d$.
Once the drive is low enough, the grain boundaries reform 
and the hysteresis is lost, as shown in Fig.~\ref{fig:3}(a).    

For $f = 3.965$, a similar sequence of phases occur with similar 
characteristics in the velocity histograms; however, 
in this case the moving grain boundary regime
persists up to higher drives. 
We find the same general features of the 
velocity histograms and dynamic phases for other fillings where grain 
boundary 
formation occurs in the pinned state. Such grain boundary states 
usually occur close to the integer fillings such as near 
$f = 3.0$ and $f = 4.0$.  

\subsection{Stripes and Symmetry Breaking Flows}

\begin{figure}
\includegraphics[width=3.5in]{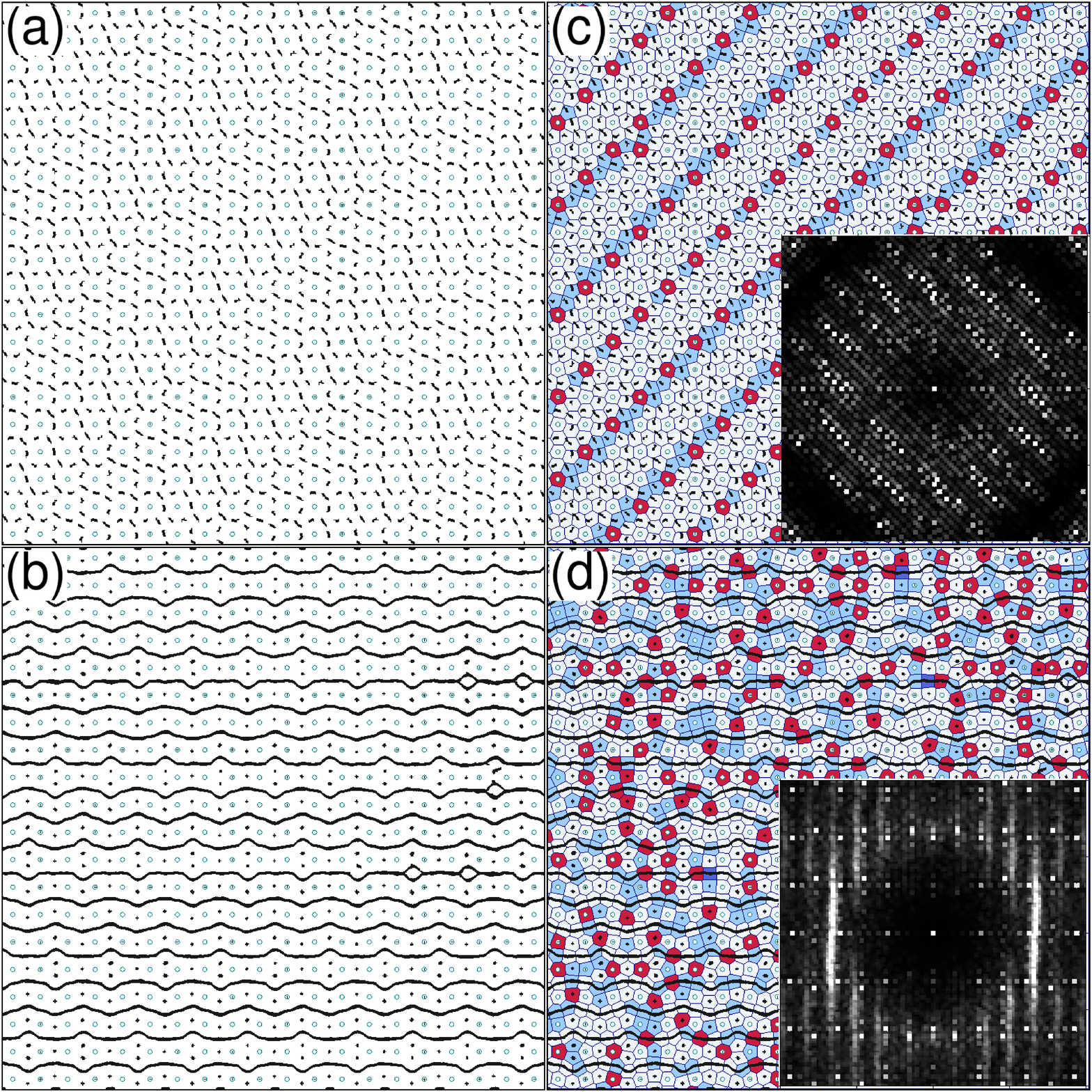}
\caption{
The dynamics at f = 4.3, where the zero drive state forms a stripe 
pattern of topological defects.
(a) The particle trajectories over time for 
the moving stripe state at $F_{d} = 0.05$ showing that the
particles are not moving along the direction of the drive but are moving in
the positive $x$ and negative $y$ directions.
(b) The particle trajectories over time for the disordered
flow state at $F_d=0.035$, indicating that the particles form 1D channels
of flow in the direction of the drive.
(c) Voronoi construction (blue lines) of 
the moving stripe state in (a). 
Inset: the structure factor $S(k)$ indicates that
the system is ordered. 
(d) Voronoi construction of 
the disordered flow state in (b).  Inset: $S(k)$ shows
that the sample has a smectic type ordering. 
The Voronoi polygons are colored according to their number of sides: 4 (dark
blue), 5 (light blue), 6 (white), and 7 (red).
}
\label{fig:4}
\end{figure}

At filling fractions where 
the topological defects form stripe structures,
the system initially depins into a moving stripe phase, 
such as shown in Fig.~\ref{fig:4}(a) for  $f = 4.3$ and 
$F_{d} = 0.05$.  
The moving state is highly ordered, as indicated by
the structure factor $S(k)$ plotted in the inset of Fig.~\ref{fig:4}(c).
At higher drives there is a transition to a state where the stripe structure 
breaks down and the   
system becomes more disordered, 
as shown in Fig.~\ref{fig:4}(c) for $F_{d} =0.35$.
In the inset of Fig.~\ref{fig:4}(c), the anisotropic features in
$S(k)$ reveal that the system is more 
disordered with smectic ordering along the $x$-direction.

In the moving stripe phase, the particles do not flow along the
$x$ direction but instead move at an angle to the driving direction, as
shown in Fig.~\ref{fig:4}(a) for the state in Fig.~\ref{fig:4}(c).
Here the particles travel
in both the positive $x$ and negative $y$ directions, 
while at $F_{d} = 0.35$, Fig.~4(b) indicates that 
the flow is now strictly in the $x$-direction. 
In the moving smectic 
phase at $F_{d} = 0.35$, the particles at the pinning sites 
and some of the interstitial particles are immobile,
and the moving interstitial
particles form channels of 1D flow.
The number of particles can vary in each of the moving rows,
causing most of the topological defects shown in 
Fig.~\ref{fig:4}(d) to align with the direction of drive and producing the
anisotropic or smectic type ordering shown in $S(k)$ in 
the inset of Fig.~\ref{fig:4}(d). 
Smectic type flows of particles in  periodic 
pinning array structures have been observed 
in simulations of vortices and colloidal particles moving over 
quasiperiodic arrays \cite{41,12} as
well as for the incommensurate flow in periodic arrays \cite{60}.           

\begin{figure}
\includegraphics[width=3.5in]{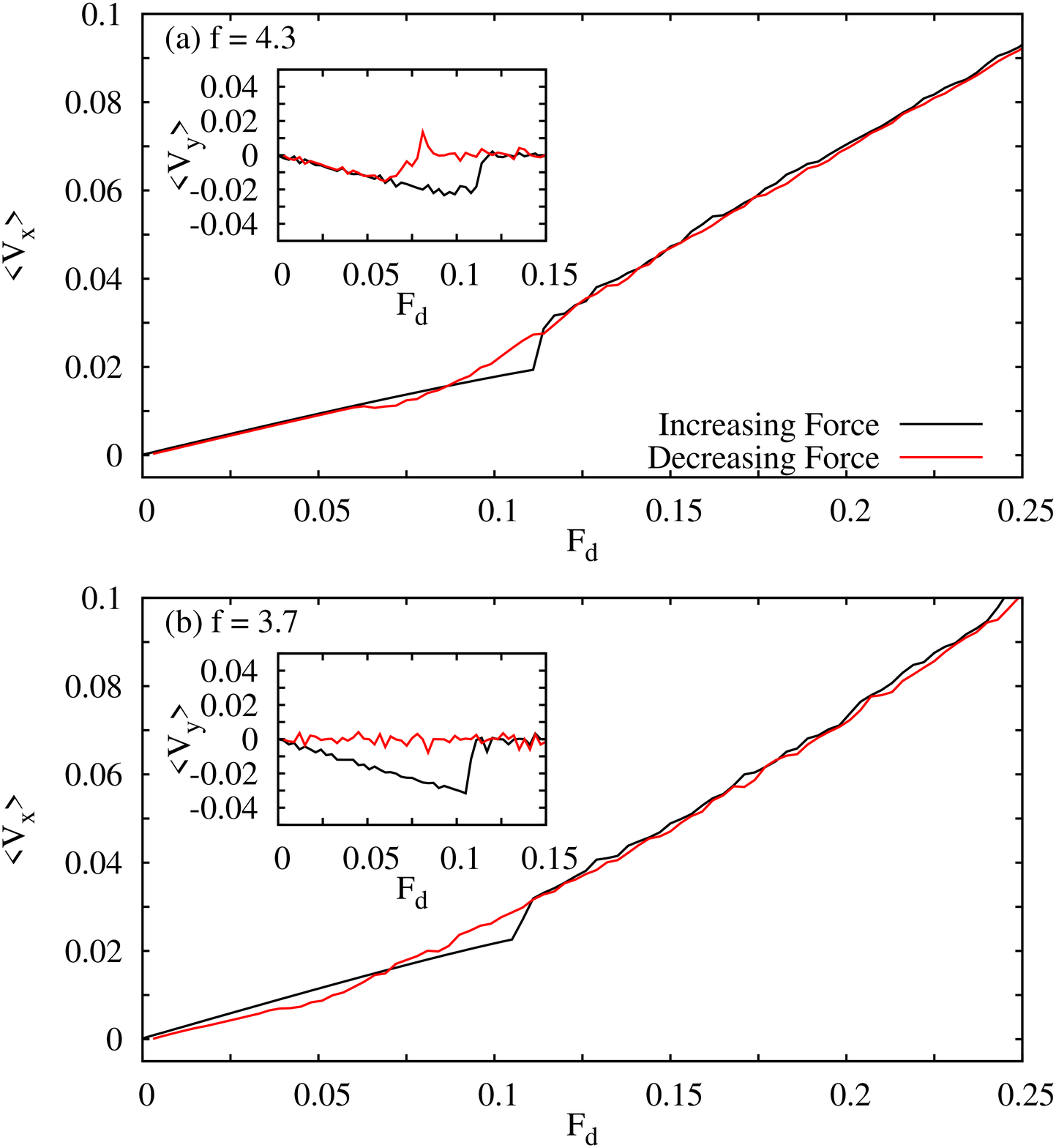}
\caption{
The $\langle V_x\rangle$ vs $F_d$ curves for increasing (black) and 
decreasing (red) sweeps of $F_d$ at fillings that form pinned stripe states.
(a) At $f= 4.3$, there is a jump up in $\langle V_{x} \rangle$ near 
$F_{d} = 0.1$ at the transition from stripe flow to smectic flow.
Inset: the corresponding 
$\langle V_{y}\rangle$ vs $F_d$ shows that in the stripe flow regime, 
the particles are also flowing in the negative $y$-direction, but this motion
is lost at the stripe-smectic flow transition.
There is hysteresis across the stripe-smectic flow transition.
(b) $\langle V_x\rangle$ vs $F_d$ for $f = 3.7$. 
Here the smectic flow is stable all the way down to $F_{d} = 0.0$.    
Inset: The corresponding $\langle V_y\rangle$ vs $F_d$.
}
\label{fig:5}
\end{figure}

The symmetry breaking flow produces signatures in the velocity force
curves $\langle V_x\rangle$ vs $F_d$, as shown
in Fig.~\ref{fig:5}(a,b) for two states that form pinned
periodic stripes, $f = 4.3$ and $f=3.7$.
Here the black line is for the initial upward sweep of $F_d$, and the red
line is for sweeping $F_d$ back down to zero.
Neither of these fillings have a finite depinning threshold within our
resolution; however, both have a two-step velocity response, which for 
$f = 4.3$ in Fig.~\ref{fig:5}(a) is associated with a jump up in 
$\langle V_{x}\rangle$ just above $F_{d} = 0.1$. 
In the inset of Fig.~\ref{fig:5}(a) 
we plot the corresponding $\langle V_{y}\rangle$ vs $F_{d}$ where we find
a linear increase in the negative $y$-direction velocity
followed by a sharp jump into a fluctuating state with
$\langle V_{y}\rangle = 0$ at the same value of $F_d$ where
a jump in $\langle V_x\rangle$ appears in the main panel.
This feature marks the transition from 
symmetry breaking flow of stripes to 
the 1D winding smectic flow. 
The velocity-force curves in Fig.~\ref{fig:5}(a) and its inset also show 
hysteresis across the stripe-smectic flow boundary, indicating that 
when the external drive is reversed, the system
can remain in the smectic flow state down to a drive 
lower than that at which the state first appeared during the
increasing sweep of $F_d$.
For $f = 3.7$, Fig.~\ref{fig:5}(b) shows that
a similar set of dynamics occurs; however, in this case the
smectic flow persists for the decreasing sweep of
$F_d$ all the way down to $F_{d} = 0.0$.     
In general, we observe symmetry
breaking flow in regimes where stripe domain wall patterns form. 
Symmetry breaking flows
have also been found for driven vortex systems 
when the vortices form effective composite objects such as dimers that
possess
an orientational degree of freedom \cite{51}. 
This is different from the situation we observe 
here, where the symmetry breaking is 
a result of the large scale symmetric patterns that form.
Symmetry breaking flow has also been observed in simulations 
of colloidal particles moving over egg-carton arrays in the
weak substrate limit when the particles
form a triangular lattice that does not have the same orientation as the
substrate lattice \cite{16}.
In that case the entire lattice flows elastically. This
differs from 
the flow shown in Fig.~\ref{fig:4}, 
where the motion is confined only to the grain boundaries and does not involve
all the particles in the system.

\begin{figure*}
\includegraphics[width=5.0in]{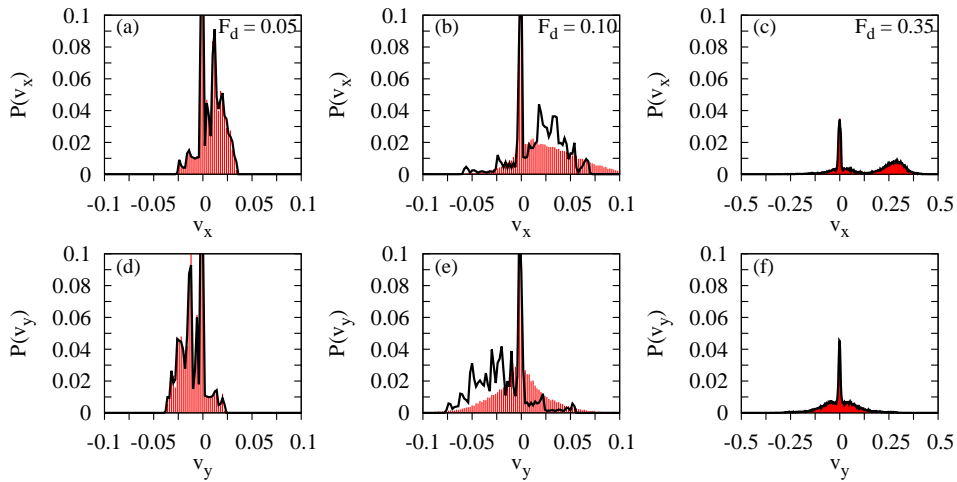}
\caption{
The velocity histograms for $f = 4.3$. 
(a,b,c) $P(v_{x})$; (d,e,f) $P(v_{y})$. 
(a,d) $F_{d} = 0.05$. (b,e) $F_{d} = 0.1$. (c,e) $F_{d} = 0.35$.     
Black lines: Increasing sweep of $F_d$.  Filled red curves: Decreasing
sweep of $F_d$.
}
\label{fig:6}
\end{figure*}

In Fig.~\ref{fig:6}(a,d) we show the velocity histograms at 
$f = 4.3$ and $F_{d} = 0.05$. 
Here $P(v_{x})$ has a positive average value with several peaks,
while the corresponding $P(v_{y})$ 
has a net negative value due to the symmetry breaking flow along the negative 
$y$-direction. 
The additional peaks in the distributions arise due to the highly ordered
flow that occurs in this phase.  The particles
move in a synchronized fashion, producing a 
periodic velocity time series.      
For $F_{d} = 0.10$ in Fig.~\ref{fig:6}(b,e), 
the stripe flow occurs for the initial ramp up of the external drive; however, 
during the ramp down, the system remains in a smectic flow regime 
so that  $\langle V_{y}\rangle  = 0.0$, as seen in the
symmetric distribution of $P(v_{y})$ in Fig.~\ref{fig:6}(e). 
At $F_{d} = 0.35$, $P(v_{x})$ in Fig.~6(c) has two prominent peaks from the
pinned and flowing particles, but
the additional smaller scale peaks that appeared for the
symmetry breaking flow are lost 
since in the smectic flow regime, the different channels move at 
different velocities, and the resulting more disordered flow
smears out the velocity distributions.  
The corresponding $P(v_{y})$ has a symmetric profile, 
indicating that the flow is oriented in the $x$-direction.
The two satellite peaks in $P(v_y)$ are due to the semi-periodic oscillations
of a number of the channels, as shown in the trajectory images 
in Fig.~\ref{fig:4}(b).  
We observe a similar set of velocity distributions at $f = 3.7$ 
as well as at other fillings where the stripe state appears.
At $f = 4.5$, the static phase does not form stripes but 
instead organizes into a checkerboard pattern \cite{1}. 
When this pattern is driven, it 
transforms into a stripe phase and shows dynamics similar to those
described for the $f=4.3$ system.

\section{Flow from $4.6 < f \leq 5.0$}

\begin{figure}
\includegraphics[width=3.5in]{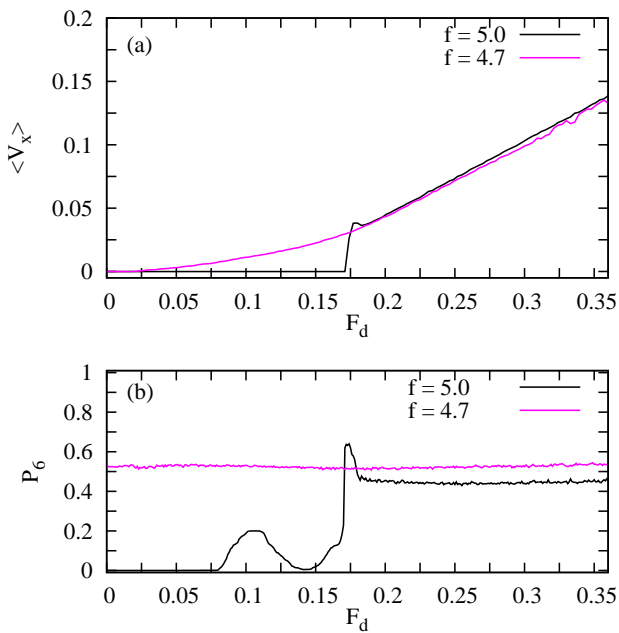}
\caption{
(a) $\langle V_x\rangle$ vs $F_d$ for a filling of
$f = 4.7$ (black) and $f = 5.0$ (purple). 
(b) The corresponding $P_{6}$ vs $F_{d}$. 
}
\label{fig:7}
\end{figure}

\begin{figure*}
\includegraphics[width=7.0in]{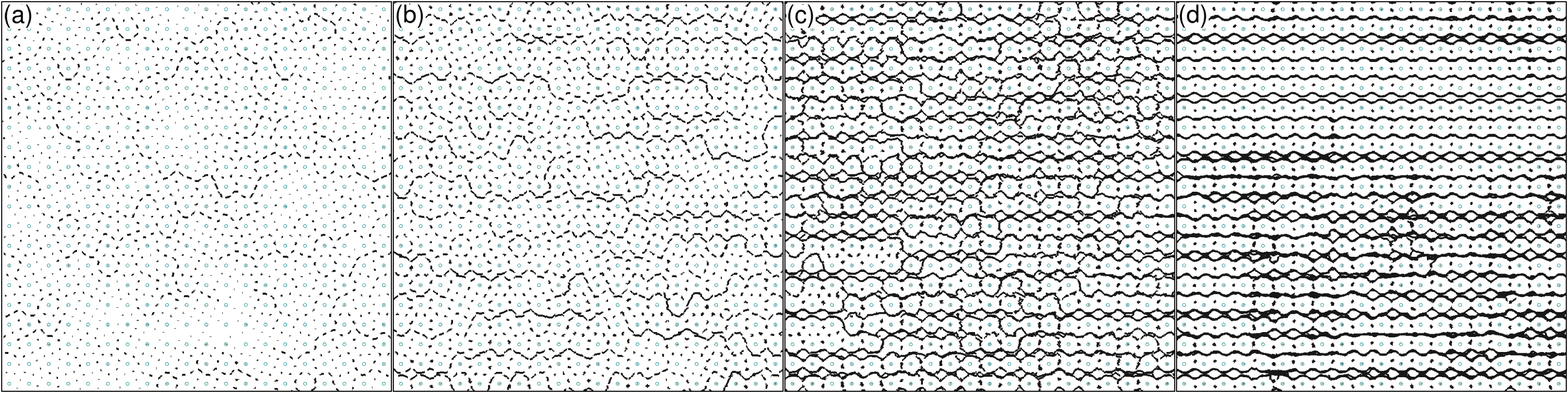}
\caption
{
The particle trajectories over a fixed time interval
for the $f = 4.7$ state. (a) $F_{d} = 0.05$ (b) $F_{d} =0.1$, (c) $F_{d} = 0.2$ 
and
(d) $F_{d} = 0.35$. 
}
\label{fig:8}
\end{figure*}

In the range $ 4.6 < f < 5.0$, the static ordered 
stripe states are replaced by disordered patch regimes.
The patches consist of regions
of square ordering that
grow in extent with increasing $f$ until the entire system forms a
square lattice at $f=5.0$.
\cite{1}. 
The depinning threshold is finite in the patch regime;
however, 
the velocity response is smooth as shown in 
Fig. \ref{fig:7}(a) where we plot
$\langle V_{x}\rangle$ vs $F_{d}$ for $f = 4.7$.
At $f = 5.0$, Fig.~\ref{fig:7}(a) shows that there is a sharp 
finite depinning threshold,  
while at $f = 4.7$ 
the depinning threshold occurs at a lower value of $F_{d} = 0.025$ and
the response above depinning shows a smooth increase. 
The particle trajectories 
above depinning for $f = 4.7$, shown in Fig.~\ref{fig:8}(a) 
for $F_d=0.05$, indicate that
disordered winding channels of flow form,
composed of localized 
soliton type pulses randomly distributed throughout the system. 
At $F_{d} = 0.1$, shown in Fig.~\ref{fig:8}(b), these
disordered regions
start to proliferate. 
At the higher drive $F_{d} = 0.2$, illustrated in Fig.~\ref{fig:8}(c), 
the flow is more strongly disordered, 
while at $F_{d} = 0.35$, shown in Fig.~\ref{fig:8}(d), 
the flow starts to become more ordered
and is confined into 1D winding channels in the interstitial regions.  
Figure \ref{fig:7}(b)  shows that $P_{6}$ vs $F_{d}$  for $f = 4.7$ 
undergoes very little change over the range of drives examined. 
In contrast, for $f = 5.0$, $P_{6}$ starts off
at zero since the system exhibits square ordering. 
Within the pinned regime, a portion of the 
particles are not directly trapped by the pinning sites so their positions
can distort under the external drive, causing the
initial pinned 
square order to transform under the external drive 
even in the absence of particle flow.   
This produces a peak in $P_{6}$ near $F_{d} = 0.1$. 
As $F_d$ continues to increase,
the system disorders, producing a drop 
in $P_{6}$ below the depinning threshold.  For higher drives
$P_6$ reaches a steady state value.
For fillings above $f = 5.0$, the
depinning threshold decreases and there is a two stage depinning response.

At $f = 4.7$, there is no hysteresis in the response to a drive.  
The velocity force characteristics for $f = 4.7$ 
in Fig.~\ref{fig:7}(a) can be fit to the 
power-law form $V \propto (F_{d} - F_{c})^\beta$,
where $F_{c}$ is the critical depinning force. 
We find $1.5 < \beta < 2.0$,  consistent with the values
obtained for disordered flows in the depinning of colloidal systems 
driven over random substrates \cite{54} and for vortices driven over
random pinning \cite{55}. 
This indicates that although the system has an underlying periodic substrate, 
there are certain fillings where the strong structural disorder caused by 
frustration effects can produce dynamics that resemble those found for systems
with random substrates.

\begin{figure}
\includegraphics[width=3.5in]{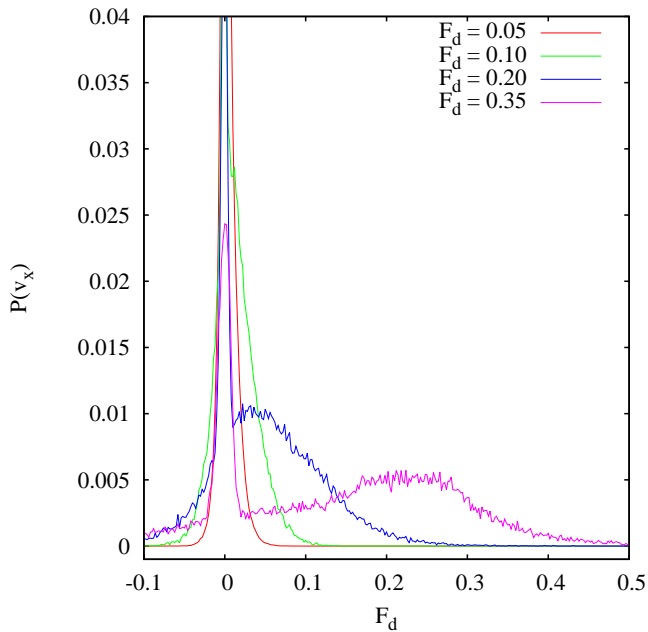}
\caption{
The velocity histograms $P(v_{x})$ for $f =4.7$ at 
$F_{d} = 0.05$ (red), 0.1 (green), 0.2 (blue), and $0.35$ (purple).
}
\label{fig:9}
\end{figure}

The velocity histograms $P(v_{x})$ at $f = 4.7$ plotted in Fig.~\ref{fig:9} 
for different values of $F_d$
show that there is no gap between the zero velocity component peak
and the higher velocity peak
in the disordered flow regime, even at the higher drive of $F_{d} = 0.35$. 
The peak at higher $v_{x}$ is
strongly smeared. 
Simulations of vortices moving over random substrates produced 
a series of velocity histograms for increasing drive with similar 
characteristics \cite{53}. 
For the filling at $f = 4.7$, the initial motion near $F_{d} = 0.05$ has a
soliton or crinkle type form where pulses move through the system and individual
particles move only about a lattice constant
each time the pulse passes by.  Unlike the more ordered fillings,
the pulse motion is not strictly confined to 1D
but can show considerable transverse mobility as well. 
The system gradually transitions to 
a state with a combination of soliton type motion 
and continuous motion near $F_{d} = 0.2$, while at $F_{d} = 0.35$
the motion is mostly continuous.          

\section{Flow Near $f = 3.0$ and $2.0$}

\begin{figure}
\includegraphics[width=3.5in]{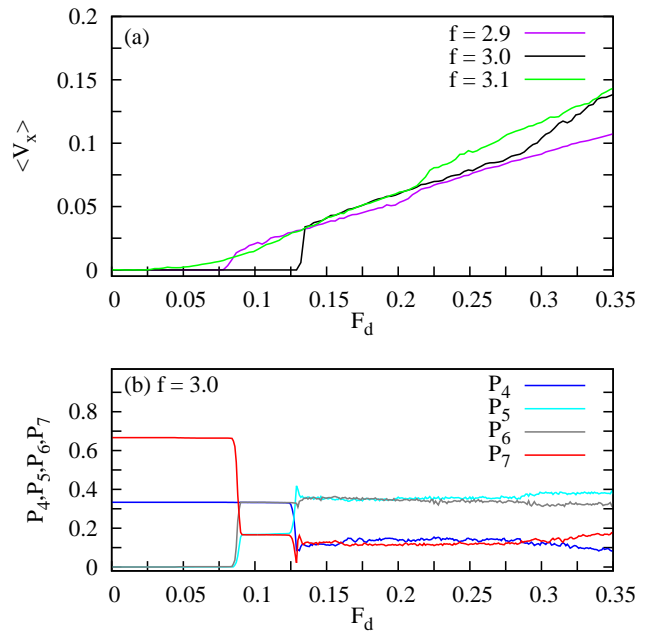}
\caption{
(a) $\langle V_{x}\rangle$ vs $F_{d}$ for $f = 3.0$ 
(highest depinning threshold), $f = 2.9$ 
(middle depinning threshold), and
$f = 3.1$ (lowest depinning threshold). 
(b) $P_{4}$ (dark blue), $P_{5}$ (light blue), $P_{6}$ (gray), 
and $P_{7}$ (red) vs $F_{d}$ 
for the $f = 3.0$ case. Here there
is a structural transition within the pinned phase to an anisotropic lattice. 
}
\label{fig:10}
\end{figure}

\begin{figure}
\includegraphics[width=3.5in]{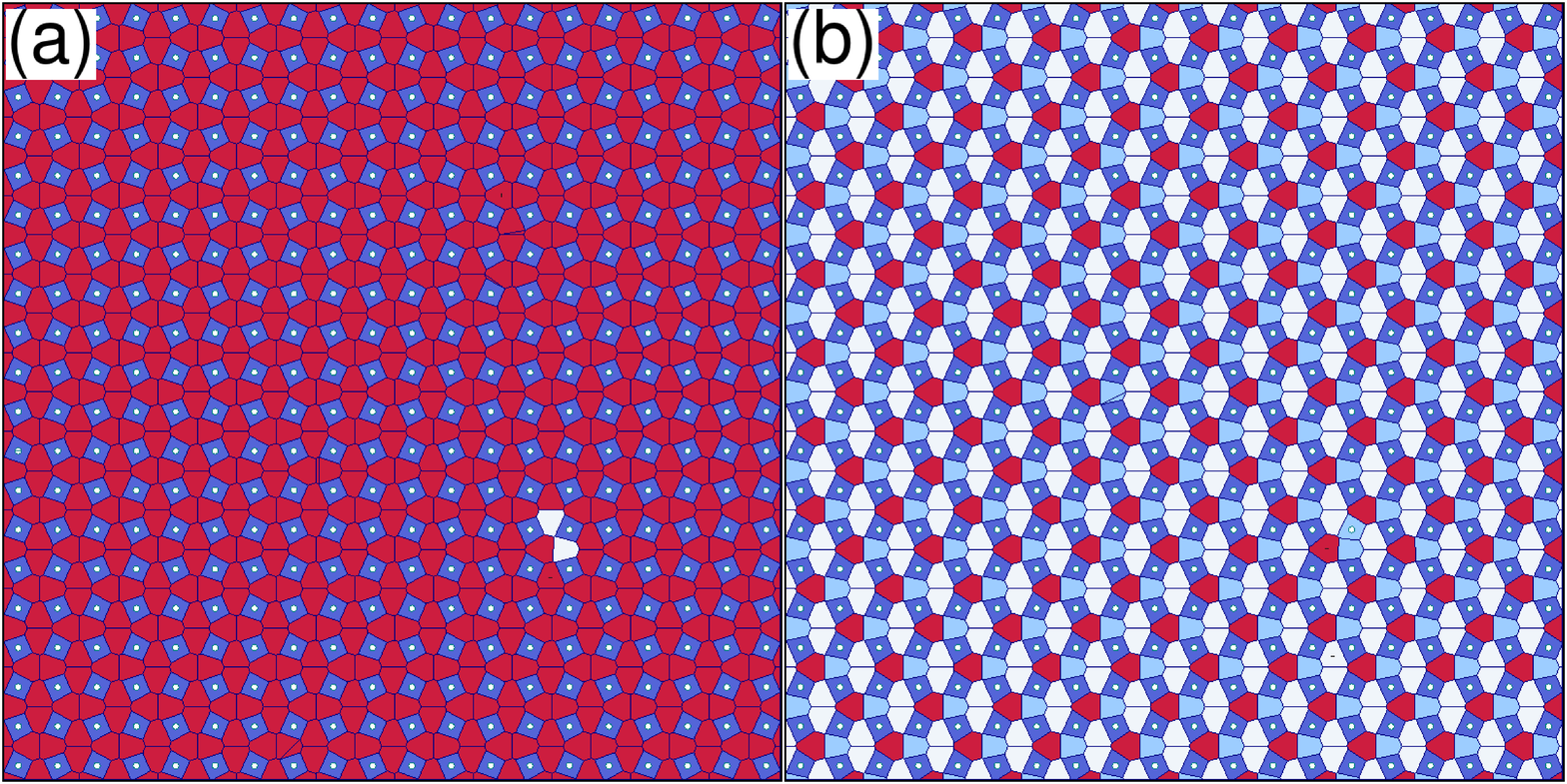}
\caption{
The Voronoi construction (blue lines) for the $f = 3.0$ state at 
(a) $F_{d} =0.05$, at the moment when the ground state first begins to
distort (indicated by the two white polygons), and
(b) $F_{d} = 0.1$ showing the transformed pinned structure.
The Voronoi polygons are colored according to their number of sides:
4 (dark blue), 5 (light blue), 6 (white), and 7 (red). 
}
\label{fig:11}
\end{figure}

For fillings just above and below $f = 3.0$, the system
forms grain boundaries and the dynamics are similar to those
observed for the grain boundary forming
states near $f = 4.0$.
In Fig.~\ref{fig:10}(a) we plot 
$\langle V_{x}\rangle$  vs $F_{d}$ for $f = 2.9$, $3.0$, and $3.1$. 
Here the  maximum in the depinning threshold
occurs for $f = 3.0$
near $F_{d} = 0.125$,
with a lower depinning threshold for $f=2.9$ and an even lower threshold
for $f=3.1.$ 
In all cases there is a 
multiple step depinning process. 
The initial motion for $f = 3.1$ and $f=2.9$ occurs
via the depinning of the grain boundaries.
In Fig.~\ref{fig:10}(b) we plot 
$P_{4}$, $P_{5}$, $P_{6}$, and $P_{7}$ for the $f = 3.0$ state. 
Here the ordered ground state persists up to $F_{d} = 0.075$, at 
which point the interstitially pinned particles have their positions distorted
strongly enough that a new pinned structure forms.
This structural transformation within the pinned state is 
similar to the structural transition discussed earlier for 
the $f = 5.0$ sample.
Just above $F_{d} = 0.125$, depinning occurs and is accompanied by a 
small jump in $P_{n}$, with $n=4$, 5, 6, 7.
The structural transformation is illustrated in Fig.~\ref{fig:11}(a) where we
plot the Voronoi construction at $F_{d} = 0.05$ in the ground state pinned
structure at the moment when it first begins to distort noticeably,
while in Fig.~\ref{fig:11}(b) we show the anisotropic
pinned structure at $F_{d} = 0.10$. 

\begin{figure}
\includegraphics[width=3.5in]{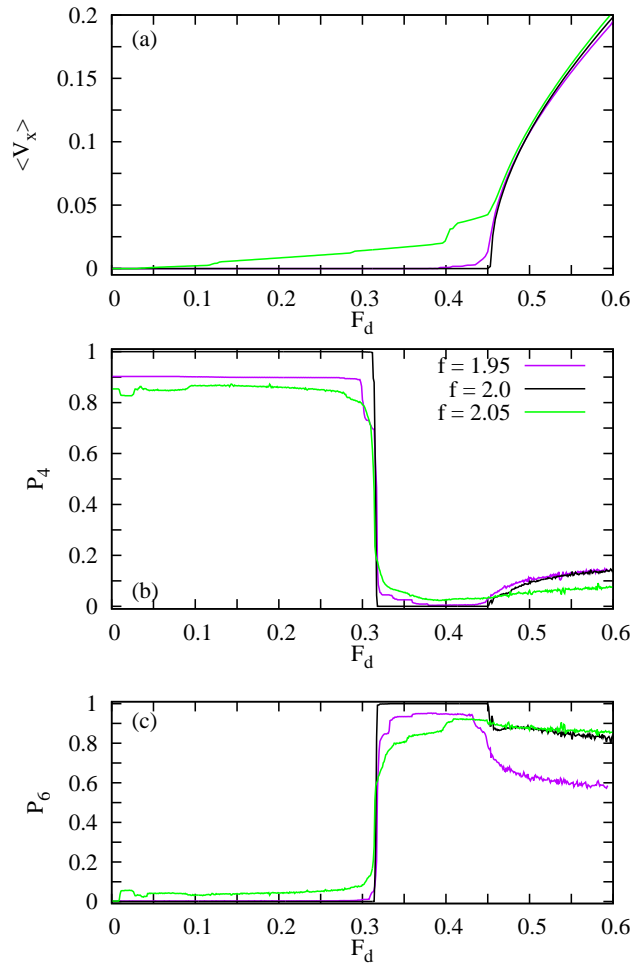}
\caption{ 
(a) $\langle V_{x}\rangle$ vs $F_{d}$ for $f= 2.0$ 
(highest depinning threshold) $f = 1.95$ (middle depinning threshold),  
and $f = 2.05$ (lowest depinning threshold). 
(b) The corresponding $P_{4}$ vs $F_d$. (c) The corresponding $P_{6}$ vs $F_d$.
}
\label{fig:12}
\end{figure}

At $f = 2.0$ the system forms a square lattice, while at 
$f = 1.95$ and $f = 2.05$ the
incommensurate configurations do not form grain boundaries 
but rather isolated islands similar to the
fillings just above and below $f = 1.0$. 
In Fig.~12(a) we plot $V_{x}$ vs $F_{d}$ for $f=1.95$, 2.0, and 2.05.
The highest depinning threshold occurs for $f = 2.0$, where 
the depinning occurs via the continuous flow of the 
interstitial particles while the particles
at the pinning sites remain pinned. 
At drives higher than those shown in the figure, a second depinning transition
occurs when the particles at the pinning sites become mobile.
At $ f= 2.05$, the interstitial depinning occurs
in two steps.  Above the first step, we find soliton type motion of the
extra interstitial particles in the square lattice ground state, while
above the second step we find the same type of depinning that occurs at $f=2.0$
where all the interstitial particles move.
For $ f = 1.95$, a similar scenario occurs; however, the depinning threshold 
is higher than for $f=2.05$ since for $f=1.95$ the initial depinning is of
vacancies or anti-kinks which have a higher depinning threshold than the 
kinks.       
In Fig.~\ref{fig:12}(b,c) we plot $P_4$ and $P_6$, respectively, versus $F_d$
for the same three fillings.
At $f = 2.0$ we observe a transition in the pinned state
from a square lattice to a 
disordered lattice with sixfold ordering due to the shifting of the
interstitial particles under the applied drive.
This is followed by a jump into a more disordered flowing state. 
A similar trend occurs for the incommensurate cases with additional disorder.

\section{Summary}

We have investigated the sliding dynamics for colloidal particles on 
periodic two dimensional muffin-tin type pinning arrays. 
In the non-driven regime, this system was previously shown to exhibit
pattern formation in the form of domain walls, stripes, and disordered phases
at incommensurate fillings, particularly in the range $ 4.0 < f < 5.0$. 
Here we find that a rich variety of distinct dynamical phases occur in
this system
including domain wall dynamics as well as disordered and continuous flow phases 
associated with characteristic velocity distributions and
structural order. 
Transitions between dynamic phases produce distinct features in the velocity
force curves, velocity histograms, particle trajectories, and 
structural ordering. 
The system is most strongly pinned
at the commensurate fillings, while at the incommensurate fillings
the system is weakly pinned and undergoes multiple depinning transitions. 
The initial depinning occurs via soliton type motion or by the motion of 
domain walls.
In the domain wall regime, when the domain wall depins, individual particles
only move about a lattice constant each time the domain wall
moves past.
For increasing drive the domain walls start to fluctuate, 
and at higher drives there is a transition to the continuous flow of
interstitial particles when the domain wall structure breaks apart. 
In the regime where the domain walls form stripes,
the particles move at an angle with respect to the drive above
depinning, while at higher drives the stripes break apart and the 
particle motion is in the direction of drive.  
We also show that even within the pinned regimes, the external drive can 
induce structural transitions
in the system since the interstitially pinned particles can have their
positions distorted relative to the particles trapped by pinning sites.
Our results on how the motion of domain walls or incommensurations
leads to different dynamical responses should be general to other 
systems exhibiting commensurate-incommensurate transitions, 
such as friction on incommensurate surfaces where
domain walls are present, and the flow of vortices in type-II superconductors.  

\acknowledgments
This work was carried out under the auspices of the 
NNSA of the 
U.S. DoE
at 
LANL
under Contract No.
DE-AC52-06NA25396.
D.M. and J.A. received support from the ASC Summer Workshop program at LANL.

\end{document}